\documentclass[aps,prl,twocolumn,showpacs,groupedaddress,floatfix]{revtex4}  % for review and submission
\usepackage{graphicx}  % needed for figures
\usepackage{dcolumn}   % needed for some tables
\usepackage{bm}        % for math
\usepackage{amssymb}   % for math

\usepackage{color}
\usepackage{ulem}

\begin{document}
% Version 13 - 4th iteration post collaboration review (new author list)
% the following information is for internal review, please remove them for submission
%\noindent\parbox{\textwidth}{Version 8, \today \hfill Comments to {\tt d0-run2eb-004@fnal.gov}}
%\parbox{\textwidth}{Authors: H. Dong, J. Heinmiller, J. Hobbs, A. Stone and N. Varelas \hfill by Mar. 26, 2006}
%\parbox{\textwidth}{To be submitted to Phys. Lett. B \hfill}
%\noindent\parbox{\textwidth}{ Version 10, \today \hfill Post collaboration
%review}
%\hrule width \textwidth\vskip 1truemm
\hspace{5.2in} \mbox{FERMILAB-PUB-07/076-E}
%\leftline{}
%\leftline{}
%\leftline{}
%\rightline {\mbox{Fermilab-Pub-xx/xxx-E}}
%\rightline {}
\def\et{$E_T$}                          %ET
\def\et{$p_T$}                          %PT
\def\D0{D\O}                            %D0
\def\ppbar{p\overline{p}}            %ppbar
\def\ttbar{t\overline{t}}            %ttbar
\def\bbbar{b\overline{b}}            %bbbar
\def\pbarp{\overline{p}p}            %pbarp
\def\etal{{\sl et al.\ }}
\def\Zll{Z\rightarrow \ell^+\ell^-}     % Z->ll
\def\Zee{Z\rightarrow ee}           % Z->ee
\def\Zmm{Z\rightarrow \mu^+\mu^-}        % Z->mumu
\def\ee{ee}
\def\mm{\mu\mu}
\newcommand{\etadet}{\ensuremath{\eta_{\mrm det}}}
\newcommand{\met}       {\mbox{$\not\!\!E_T$}}
\newcommand{\mrm}{\mathrm}
\newcommand{\etopt}{\ensuremath{E_T/p_T}}
\newcommand{\wenu}{\ensuremath{W{\rightarrow}e^\pm\nu}}
\newcommand{\zee}{\ensurematch{Z \rightarrow ee}}
\newcommand{\pythia}    {{\sc pythia}}
\newcommand{\alpgen}    {{\sc alpgen}}
\newcommand{\evtgen}    {{\sc evtgen}}
\newcommand{\geant}    {{\sc geant}}
\newcommand{\cteq}    {{\sc cteq5l}}
\newcommand{\cteqSixL}    {{\sc cteq6l}}
\newcommand{\cteqm}   {{\sc cteq6m}}
\newcommand{\mcfm}   {{\sc mcfm}}

\title{Search for a Higgs boson produced in association with a $Z$
boson in $\ppbar$ collisions 
% at $\sqrt{s}$ = 1.96 TeV (Ferbelization)
}

% LIST_OF_AUTHORS_R2.TEX                4/11/07             
%
\author{                                                                      
%% names begin here                                                           
V.M.~Abazov,$^{35}$                                                           
B.~Abbott,$^{75}$                                                             
M.~Abolins,$^{65}$                                                            
B.S.~Acharya,$^{28}$                                                          
M.~Adams,$^{51}$                                                              
T.~Adams,$^{49}$                                                              
E.~Aguilo,$^{5}$                                                              
S.H.~Ahn,$^{30}$                                                              
M.~Ahsan,$^{59}$                                                              
G.D.~Alexeev,$^{35}$
% ...
\\                                                                            
\vskip 0.30cm                                                                 
\centerline{(D\O\ Collaboration)}                                             
\vskip 0.30cm                                                                 
}                                                                             
\affiliation{                                                                 
\centerline{$^{1}$Universidad de Buenos Aires, Buenos Aires, Argentina}       
\centerline{$^{2}$LAFEX, Centro Brasileiro de Pesquisas F{\'\i}sicas,         
                  Rio de Janeiro, Brazil}                                     
\centerline{$^{3}$Universidade do Estado do Rio de Janeiro,                   
                  Rio de Janeiro, Brazil}                                     
\centerline{$^{4}$Instituto de F\'{\i}sica Te\'orica, Universidade            
                  Estadual Paulista, S\~ao Paulo, Brazil}                     
\centerline{$^{5}$University of Alberta, Edmonton, Alberta, Canada,           
                  Simon Fraser University, Burnaby, British Columbia, Canada,}
\centerline{York University, Toronto, Ontario, Canada, and                    
                  McGill University, Montreal, Quebec, Canada}                
\centerline{$^{6}$University of Science and Technology of China, Hefei,       
                  People's Republic of China}                                 
\centerline{$^{7}$Universidad de los Andes, Bogot\'{a}, Colombia}             
\centerline{$^{8}$Center for Particle Physics, Charles University,            
                  Prague, Czech Republic}                                     
\centerline{$^{9}$Czech Technical University, Prague, Czech Republic}         
\centerline{$^{10}$Center for Particle Physics, Institute of Physics,         
                   Academy of Sciences of the Czech Republic,                 
                   Prague, Czech Republic}                                    
\centerline{$^{11}$Universidad San Francisco de Quito, Quito, Ecuador}        
\centerline{$^{12}$Laboratoire de Physique Corpusculaire, IN2P3-CNRS,         
                   Universit\'e Blaise Pascal, Clermont-Ferrand, France}      
\centerline{$^{13}$Laboratoire de Physique Subatomique et de Cosmologie,      
                   IN2P3-CNRS, Universite de Grenoble 1, Grenoble, France}    
\centerline{$^{14}$CPPM, IN2P3-CNRS, Universit\'e de la M\'editerran\'ee,     
                   Marseille, France}                                         
\centerline{$^{15}$Laboratoire de l'Acc\'el\'erateur Lin\'eaire,              
                   IN2P3-CNRS et Universit\'e Paris-Sud, Orsay, France}       
\centerline{$^{16}$LPNHE, IN2P3-CNRS, Universit\'es Paris VI and VII,         
                   Paris, France}                                             
\centerline{$^{17}$DAPNIA/Service de Physique des Particules, CEA, Saclay,    
                   France}                                                    
\centerline{$^{18}$IPHC, Universit\'e Louis Pasteur et Universit\'e           
                   de Haute Alsace, CNRS, IN2P3, Strasbourg, France}          
\centerline{$^{19}$IPNL, Universit\'e Lyon 1, CNRS/IN2P3, Villeurbanne, France
                   and Universit\'e de Lyon, Lyon, France}                    
\centerline{$^{20}$III. Physikalisches Institut A, RWTH Aachen,               
                   Aachen, Germany}                                           
\centerline{$^{21}$Physikalisches Institut, Universit{\"a}t Bonn,             
                   Bonn, Germany}                                             
\centerline{$^{22}$Physikalisches Institut, Universit{\"a}t Freiburg,         
                   Freiburg, Germany}                                         
\centerline{$^{23}$Institut f{\"u}r Physik, Universit{\"a}t Mainz,            
                   Mainz, Germany}                                            
\centerline{$^{24}$Ludwig-Maximilians-Universit{\"a}t M{\"u}nchen,            
                   M{\"u}nchen, Germany}                                      
\centerline{$^{25}$Fachbereich Physik, University of Wuppertal,               
                   Wuppertal, Germany}                                        
\centerline{$^{26}$Panjab University, Chandigarh, India}                      
\centerline{$^{27}$Delhi University, Delhi, India}                            
\centerline{$^{28}$Tata Institute of Fundamental Research, Mumbai, India}     
\centerline{$^{29}$University College Dublin, Dublin, Ireland}                
\centerline{$^{30}$Korea Detector Laboratory, Korea University,               
                   Seoul, Korea}                                              
\centerline{$^{31}$SungKyunKwan University, Suwon, Korea}                     
\centerline{$^{32}$CINVESTAV, Mexico City, Mexico}                            
\centerline{$^{33}$FOM-Institute NIKHEF and University of                     
                   Amsterdam/NIKHEF, Amsterdam, The Netherlands}              
\centerline{$^{34}$Radboud University Nijmegen/NIKHEF, Nijmegen, The          
                  Netherlands}                                                
\centerline{$^{35}$Joint Institute for Nuclear Research, Dubna, Russia}       
\centerline{$^{36}$Institute for Theoretical and Experimental Physics,        
                   Moscow, Russia}                                            
\centerline{$^{37}$Moscow State University, Moscow, Russia}                   
\centerline{$^{38}$Institute for High Energy Physics, Protvino, Russia}       
\centerline{$^{39}$Petersburg Nuclear Physics Institute,                      
                   St. Petersburg, Russia}                                    
\centerline{$^{40}$Lund University, Lund, Sweden, Royal Institute of          
                   Technology and Stockholm University, Stockholm,            
                   Sweden, and}                                               
\centerline{Uppsala University, Uppsala, Sweden}                              
\centerline{$^{41}$Physik Institut der Universit{\"a}t Z{\"u}rich,            
                   Z{\"u}rich, Switzerland}                                   
\centerline{$^{42}$Lancaster University, Lancaster, United Kingdom}           
\centerline{$^{43}$Imperial College, London, United Kingdom}                  
\centerline{$^{44}$University of Manchester, Manchester, United Kingdom}      
\centerline{$^{45}$University of Arizona, Tucson, Arizona 85721, USA}         
\centerline{$^{46}$Lawrence Berkeley National Laboratory and University of    
                   California, Berkeley, California 94720, USA}               
\centerline{$^{47}$California State University, Fresno, California 93740, USA}
\centerline{$^{48}$University of California, Riverside, California 92521, USA}
\centerline{$^{49}$Florida State University, Tallahassee, Florida 32306, USA} 
\centerline{$^{50}$Fermi National Accelerator Laboratory,                     
            Batavia, Illinois 60510, USA}                                     
\centerline{$^{51}$University of Illinois at Chicago,                         
            Chicago, Illinois 60607, USA}                                     
\centerline{$^{52}$Northern Illinois University, DeKalb, Illinois 60115, USA} 
\centerline{$^{53}$Northwestern University, Evanston, Illinois 60208, USA}    
\centerline{$^{54}$Indiana University, Bloomington, Indiana 47405, USA}       
\centerline{$^{55}$University of Notre Dame, Notre Dame, Indiana 46556, USA}  
\centerline{$^{56}$Purdue University Calumet, Hammond, Indiana 46323, USA}    
\centerline{$^{57}$Iowa State University, Ames, Iowa 50011, USA}              
\centerline{$^{58}$University of Kansas, Lawrence, Kansas 66045, USA}         
\centerline{$^{59}$Kansas State University, Manhattan, Kansas 66506, USA}     
\centerline{$^{60}$Louisiana Tech University, Ruston, Louisiana 71272, USA}   
\centerline{$^{61}$University of Maryland, College Park, Maryland 20742, USA} 
\centerline{$^{62}$Boston University, Boston, Massachusetts 02215, USA}       
\centerline{$^{63}$Northeastern University, Boston, Massachusetts 02115, USA} 
\centerline{$^{64}$University of Michigan, Ann Arbor, Michigan 48109, USA}    
\centerline{$^{65}$Michigan State University,                                 
            East Lansing, Michigan 48824, USA}                                
\centerline{$^{66}$University of Mississippi,                                 
            University, Mississippi 38677, USA}                               
\centerline{$^{67}$University of Nebraska, Lincoln, Nebraska 68588, USA}      
\centerline{$^{68}$Princeton University, Princeton, New Jersey 08544, USA}    
\centerline{$^{69}$State University of New York, Buffalo, New York 14260, USA}
\centerline{$^{70}$Columbia University, New York, New York 10027, USA}        
\centerline{$^{71}$University of Rochester, Rochester, New York 14627, USA}   
\centerline{$^{72}$State University of New York,                              
            Stony Brook, New York 11794, USA}                                 
\centerline{$^{73}$Brookhaven National Laboratory, Upton, New York 11973, USA}
\centerline{$^{74}$Langston University, Langston, Oklahoma 73050, USA}        
\centerline{$^{75}$University of Oklahoma, Norman, Oklahoma 73019, USA}       
\centerline{$^{76}$Oklahoma State University, Stillwater, Oklahoma 74078, USA}
\centerline{$^{77}$Brown University, Providence, Rhode Island 02912, USA}     
\centerline{$^{78}$University of Texas, Arlington, Texas 76019, USA}          
\centerline{$^{79}$Southern Methodist University, Dallas, Texas 75275, USA}   
\centerline{$^{80}$Rice University, Houston, Texas 77005, USA}                
\centerline{$^{81}$University of Virginia, Charlottesville,                   
            Virginia 22901, USA}                                              
\centerline{$^{82}$University of Washington, Seattle, Washington 98195, USA}  
}                                                                             
%end                                                                          

\date{April 16, 2007}

\begin{abstract}
We describe a search for the standard model Higgs boson with a mass of
105~GeV/$c^2$ to 145~GeV/$c^2$ in data corresponding to an integrated
luminosity of approximately 450~pb$^{-1}$ collected with the D0 detector at the
Fermilab Tevatron $\ppbar$ collider at a center-of-mass energy of 1.96~TeV.
The Higgs boson is required to be produced in association with a $Z$ boson, and
the $Z$ boson is required to decay to either electrons or muons with the Higgs
boson decaying to a $\bbbar$ pair.  The data are well described by the expected
background, leading to 95\% confidence level cross section upper limits
$\sigma(\ppbar\rightarrow ZH)
\times B(H\rightarrow\bbbar)$ in the range of 3.1~pb to 4.4~pb.
\end{abstract}

\maketitle
%\vspace{5cm}

Over the past two decades, increasingly precise experimental results have
repeatedly validated the standard model (SM) and the relationship between gauge
invariance and the embedded coupling strengths.  For massive $W$ and $Z$
bosons, gauge invariance of the Lagrangian is preserved through the Higgs
mechanism, but the Higgs boson ($H$) has yet to be observed.
The current lower bound on the mass of the Higgs boson from direct experimental
searches is $M_H=114.4$~GeV/$c^2$ at the 95\% confidence level~\cite{b-lep}.
Searches for $\ppbar\rightarrow WH\rightarrow e(\mu)\nu
\bbbar$, $\ppbar\rightarrow WH\rightarrow WWW^*$, and $\ppbar\rightarrow
ZH\rightarrow\nu\overline{\nu}\bbbar$ have been recently
reported~\cite{b-whOld,b-hwww,b-nunu}.  The CDF collaboration has reported
results in the $\ppbar\rightarrow WH\rightarrow \ell\nu$ and $\ppbar\rightarrow
ZH\rightarrow \ell^+\ell^-\bbbar\ \ (\ell = e, \mu)$ channels with significantly
smaller data sets~\cite{b-CDF1,b-CDF2,b-CDF}.  This Letter provides the first
results from the D0 experiment of searches for a Higgs boson produced in
association with a $Z$ boson, which then decays either to an electron pair or
to a muon pair.  The Higgs is assumed to decay to a $\bbbar$ pair with a
branching fraction given by the SM.
%The branching fraction for this mode is significant for Higgs
%masses less than roughly 135~GeV/$c^2$.  
The $Z(\rightarrow\ell^+\ell^-)H$ channels reported in this letter comprise major
components of the search for a Higgs boson at the Tevatron collider.

$Z$ bosons are reconstructed and identified through pairs of isolated,
electrons or muons with large momentum components transverse to
the beam direction ($p_T$) having invariant mass consistent with that of the
$Z$ boson.  Events are required to have exactly two jets identified as arising
from $b$ quarks ($b$ jets).  The resulting data are examined for the presence
of a ($H\rightarrow b\bar{b}$) signal in the $b$-tagged dijet mass
distribution.  An efficient $b$-identification algorithm with low
misidentification rate and good dijet mass resolution are essential to enhance
signal relative to the backgrounds.  The analysis of the
dielectron~\cite{b-jhthesis} (dimuon~\cite{b-hdthesis}) channel is based on
$450\pm27$~pb$^{-1}$ ($370\pm23$~pb$^{-1}$) of data recorded by the D0 experiment between
2002 and 2004.

The D0 detector~\cite{run1det,run2det} has a central-tracking system consisting
of a silicon microstrip tracker (SMT) and a central fiber tracker (CFT), both
located within a $\approx 2$~T superconducting solenoidal magnet, with designs
optimized for tracking and vertexing covering pseudorapidities $|\eta|<3$ and
$|\eta|<2.5$, respectively ($\eta = -\ln[\tan(\theta/2)]$, with $\theta$ the
polar angle relative to the direction of the proton beam).  Central and forward
preshower detectors are positioned just outside of the superconducting coil. A
liquid-argon and uranium calorimeter has a central section (CC) covering
pseudorapidities up to $|\eta| \approx 1.1$ and two end calorimeters (EC) that
extend coverage to $|\eta|\approx 4.2$, with all three housed in separate
cryostats~\cite{run1det}. An outer muon system, covering $|\eta|<2$, consists
of a layer of tracking detectors and scintillation trigger counters in front of
1.8~T toroids, followed by two similar layers behind the
toroids~\cite{run2muon}.  Luminosity is measured using plastic scintillator
arrays placed in front of the EC cryostats~\cite{b-lumi}.

The primary background to the Higgs signal is the associated production of a
$Z$ boson with jets arising from gluon radiation, among which $Z+b\bar{b}$
production is an irreducible background.  The other background sources are
$t\bar{t}$ production, diboson ($ZZ$ and $WZ$) production, and events from
multijet production that are misidentified as containing $Z$ bosons.  The
backgrounds are grouped into two categories with the first category, called
physics backgrounds, containing events with $Z$ or $W$ bosons arising from SM
processes: inclusive $Z+b\bar{b}$ production, inclusive $Z+jj$ production in
which $j$ is a jet without $b$ flavor, $t\bar{t}$, $ZZ$, and $WZ$ events.  This
background is estimated from simulation as described below.  The
second category, called instrumental background, contains those events from
multijet production that have two jets misidentified as isolated electrons or
muons which appear to arise from the $Z$ boson decay.  This background is
modeled using control data samples and the procedure described below.

Physics backgrounds are simulated using the leading order \alpgen~\cite{b-alpgen} and
\pythia~\cite{b-pythia} event generators, with the leading order \cteq~\cite{b-cteq} used as 
parton distribution functions.  The decay and fragmentation of heavy flavor
hadrons is done via \evtgen~\cite{b-evtgen}.  The simulated events
are passed through a detailed D0 detector simulation program based on
\geant~\cite{b-geant} and are reconstructed using the same software program
used to reconstruct the collider data.  The $ZH$ signal, for a range of Higgs
masses, is also simulated using \pythia\ with the same processing as applied to
data.  Determination of the instrumental background and the normalization of
the physics backgrounds are discussed below.

Candidate $Z\rightarrow \ee$ events are selected using a combination of
single-electron triggers.  Accepted events must have two isolated
electromagnetic (EM) clusters reconstructed offline in the calorimeter.
% Start change
Isolation is defined as $I = (E_{\mathrm{total}}^{(0.4)}-E_{EM}^{(0.2)})/
E_{EM}^{(0.2)}$ in which $E_{\mathrm{total}}^{(0.4)}$ is the total
calorimeter energy within $\Delta R<0.4$ of the electron direction and
$E_{EM}^{(0.2)}$ is the energy in the electromagnetic portion of the
calorimeter within $\Delta R<0.2$ of the electron direction. Candidate
electrons must satisfy $I < 0.15$.
% end change
Each EM cluster must have $p_{T}> 20$~GeV/$c$ and either
$|\eta_{\text{det}}|<1.1$ or $1.5<|\eta_{\text{det}}|<2.5$, where
$\eta_{\text{det}}$ is the pseudorapidity measured relative to the center of
the detector, with at least one cluster satisfying $|\eta_{\text{det}}|<1.1$.
In addition, the lateral and longitudinal shower shape of the energy cluster
must be consistent with that expected of electrons. 
%start change
% as determined using a
% $\chi^2$ comparison.  Electrons are retained if
% $\chi^2<12(20)$ in the central(forward) region.  
%end change
At least one of the two EM
clusters is also required to have a reconstructed track matching the position
of the EM cluster energy.  Events with a dielectron mass of
$75<M_{ee}<105$~GeV/$c^2$ form the $Z$ boson candidate sample in the dielectron
channel.

Candidate $Z\rightarrow\mu^+\mu^-$ events are selected using a set of
single-muon triggers.  Accepted events must have two isolated muons
reconstructed offline.  The muons must have opposite charge,
$p_{T}>15$~GeV/$c$, and $|\eta|<2.0$ with muon trajectories matched to tracks
in the central tracking system (i.e., the SMT and the CFT), where the central
track must contain at least one SMT measurement.  In addition,
the central tracks are required to have a distance of closest approach to the
interaction vertex in the transverse plane smaller than 0.25~cm.  
% start deletion
Muon
isolation is based on the sum of the energy measured in the calorimeter around
the muon candidate and the sum of the $p_{T}$ of tracks within $\Delta R
=\sqrt{(\Delta\phi)^2+(\Delta\eta)^2} = 0.5$ of the muon candidate normalized
by the muon momentum. 
% end deletion
% start change
The distribution of this variable in background
multihadron events is converted to a probability distribution such that a low
probability corresponds to an isolated muon.  The product of the probabilities
for both muons in an event is computed, and the event is retained if the
product is less than 0.02.  
% end change
Accepted $Z$ boson candidates must have the opening
angle of the dimuon system in the transverse plane (azimuth) of $\Delta
\phi >0.4$, and invariant mass $65\ \mathrm{GeV}/c^2 < M_{\mu\mu} < 115$~GeV/$c^2$.
This mass range differs from that of the dielectrons because of the difference in
resolutions of electron energies and muon momenta.

After selecting the $Z$ candidate events, we define a $Z$+dijet sample which,
in addition to satisfying the $Z$ candidate selection requirements, has at
least two jets in each event.  Jets are reconstructed from energy in
calorimeter towers using the Run~II cone algorithm with $\Delta R =
0.5$~\cite{ref:jet} with towers defined as non-overlapping, adjacent regions of
the calorimeter $\Delta\eta \times \Delta\phi = 0.1 \times 0.1$ in size.  The
transverse momentum of each jet is corrected for multiple $\ppbar$
interactions, calorimeter noise, out-of-cone showering in the calorimeter, and
energy response of the calorimeter as determined from the transverse momentum
imbalance in photon+jet events~\cite{b-JES1,b-JES2}.  Only jets that pass standard
quality requirements and satisfy $p_{T} > 20$~GeV/$c$ and $|\eta|<2.5$ are used
in this analysis.  The quality requirements are based on the pattern of energy
deposition within a jet and consistency with the energy deposition measured by
the trigger system.

For the $\Zee$ channel, the normalizations of the smaller $\ttbar$,
$WZ$ and $ZZ$ backgrounds are computed using simulated events and
next-to-leading-order (NLO) cross sections.  Trigger efficiency, electron
identification (ID) efficiency and resolution correction factors are derived
from comparisons of data control samples and simulated events.  The background
contributions from $Z+jj$, $Z+bj$ and $Z+\bbbar$ processes are normalized to
the observed $Z+$dijet data yield reduced by the expected contributions from
the smaller physics and instrumental backgrounds.  The relative
fractions of the $Z+jj$, $Z+bj$ and $Z+\bbbar$ backgrounds in the $Z+$dijet
sample are determined from the acceptance and selection efficiencies multiplied
by the ratios of the NLO cross sections for these processes computed using the
\mcfm~\cite{b-mcfm} program and the next-to-leading order \cteqm~\cite{b-cteq6} 
parton distribution functions.  
For the $\Zmm$ channel, all physics backgrounds are determined
using simulated events with NLO cross sections applied.  Trigger efficiency,
muon ID efficiency, and resolution correction factors are derived from
comparison of data control samples and the simulated events.%~\cite{f-comment}.

Instrumental backgrounds in both channels are determined by fitting the
dilepton invariant mass distributions to a sum of non-$Z$ and $Z$ boson
contributions.  The $Z$ boson lineshape is modeled using a Breit-Wigner
distribution convoluted with a Gaussian representing detector resolution.  The
non-$Z$ background, consisting of a sum of events from Drell-Yan production and
instrumental background, is modeled using exponentials.  The ratio of $Z$ boson
to non-resonant Drell-Yan production is fixed by the standard model.

The (two) jets arising from Higgs boson decay should contain $b$-flavored
quarks ($b$ jets), whereas background from $Z+$jets has relatively few events
with $b$ jets.  To improve the signal-to-background ratio, two of the jets in
the events from the $Z$+dijet sample are required to exhibit properties consistent
with those of jets containing $b$ quarks.
%
%The hadronic jet
%is defined as ``taggable'' if it is associated with a cluster of tracks
%($\equiv$ track jet) within $\Delta R < 0.5$.  The track jet is found by
%applying a cone track clustering algorithm with the size $\Delta R = 0.5$.  The
%track jet is required to have at least two tracks, where the seed track must
%have $p_{T}>1$~GeV/c, and the other $p_{T}>0.5$~GeV/c.  Each track must be within
%2~cm of the interaction vertex in $z$ and have at least one SMT hit.
%
The same $b$-jet identification algorithm~\cite{b-jlip} is used for the dielectron and dimuon
samples. It is based on the finite lifetime of $b$ hadrons giving
a low probability that these tracks appear to arise from the primary vertex and
considers all central tracks associated with a jet.  A small probability
corresponds to jets with tracks with large impact parameter, as expected in $b$
hadron decays.  The efficiency for tagging a $b$ jet from Higgs decay is
approximately 50\%, determined as described in the next paragraph.
The probability of misidentifying a jet arising from a charm quark as a $b$ jet
is roughly 20\%.
%%JDH: 50% = 60% TRF x 80% taggability; c < 0.5 \times b
The probability to misidentify a jet arising from a light quark ($u$, $d$, $s$)
or gluon as a $b$ jet is roughly 4\%.  This choice of efficiency and purity
optimizes the sensitivity of the analysis.  The relatively large per-jet
light-flavor misidentification rate can be accommodated because two tagged jets
are required in each event.

For background yields determined from simulated events, the probability as a
function of jet $p_T$ and $\eta$ that a jet of a given flavor would be
identified (tagged) as a $b$ jet is applied to each jet in an event.  The
probability functions are derived from control data samples.  For jets in the
simulated events, the flavor is determined from a priori knowledge of the
parton that gives rise to the jet.  The probability of having two $b$-tagged
jets is defined by convoluting the per-jet probabilities assuming there 
are no jet-to-jet correlations introduced by the b-tag requirement.
%The tagging used in simulated
%events is thus based on measurements from the data, and it does not rely
%directly on the simulation of the tagging.  
The observed
number of $Z+2\ b$-jet events and the predicted background levels are shown in
Table~\ref{t-datavmc}.

\begin{figure}[!]
\includegraphics[scale=0.46]{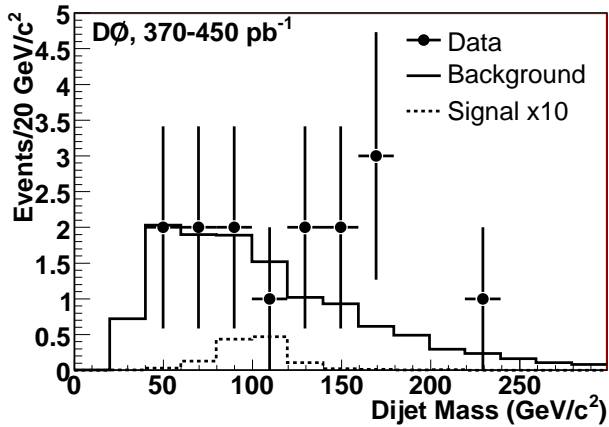}
\caption{\label{f-mass} The dijet invariant mass distribution in double--tagged
$Z$+dijet events.  The Higgs signal corresponds to $M_H = 115$~GeV/$c^2$.  (The
uncertainties are statistical only.)}
\end{figure}

The invariant mass of the two $b$ jets in the $Z+2$~$b$ jet sample is shown in
Fig.~\ref{f-mass}.  This distribution is searched for an excess of events.  The
peak position in the dijet mass spectrum is expected to be at a lower value
than the hypothetical Higgs mass because the jet energy is corrected to reflect
the energy of particles in the jet cone without a general correction for the
lower $b$ jet response compared with light jets.  If a muon is within $\Delta R
< 0.5$ of the jet, then twice the muon momentum is added to the jet
momentum.  This is an approximation to the energy of both the muon and the
accompanying neutrino.  The expected contribution from Higgs boson production
shown in Fig.~\ref{f-mass} corresponds to $M_H = 115$~GeV/$c^2$.

%%%%%%%%%%%%%%%%%%%%%%%%%%%%%%%%%%%%%%%%%%%%%%%%%%%%%%%%%%%%%%%%%%%%%%%%%%%%%%%%%%%%%%%%%%%%%%
%%%%%%%%%%%%%%%%%%%%%%%%%%%%%%%%%%%%%%%%%%%%%%%%%%%%%%%%%%%%%%%%%%%%%%%%%%%%%%%%%%%%%%%%%%%%%%
\begin{table*}[htbp]
    \caption {\label{t-datavmc}
       Number of observed and expected background events.
    }
    \begin{ruledtabular}
%    \begin{tabular}{l|lll|lll}
%        &\multicolumn{3}{c|}{$Z+\geq 2$~jets}
    \begin{tabular}{lcccrccc}
        &\multicolumn{3}{c}{$Z+\geq 2$~jets}
        & \hphantom{123}
        &\multicolumn{3}{c}{2$~b$~tags} \\
	 Final state
        &$Z\rightarrow ee$ &$Z\rightarrow \mu^{+}\mu^{-}$
        &$Z\rightarrow \ell^{+}\ell^{-}$ 
        &&$Z\rightarrow ee$ &$Z\rightarrow \mu^{+}\mu^{-}$
        &$Z\rightarrow \ell^{+}\ell^{-}$ \\
    \hline
    $Zbb$        &9.1  & 8.3  & 17.4 && 2.0  & 1.3 & 3.3  \\
    $Zjj$        &414  & 437  & 851  && 1.2  & 2.6 & 3.8  \\
    $t\bar{t}$   &2.7  & 9.6  & 12.3 && 0.80 & 3.1 & 3.9 \\
    $ZZ+WZ$      &9.2  & 21.4 & 30.6 && 0.32 & 0.42& 0.74 \\
    Instrumental &28.0 & 16.1 & 44.1 && 0.18 & 0.41& 0.59 \\
    \hline
    Total background 
               &463  & 493  &  956 && 4.5  & 7.8  & 12.3 \\
    \hline
    Observed events &463 &545 &1008 &&5 &10 &15 \\
    \end{tabular}
   \end{ruledtabular}
\end{table*}

Systematic uncertainties for signal and background arise from a variety of
sources, including uncertainties on the trigger efficiency, on
the corrections for differences between data and simulation for lepton
reconstruction and identification efficiencies, lepton energy resolution, jet
reconstruction efficiencies and energy determination, $b$-identification
efficiency, uncertainties from theory and parton distribution functions for
cross sections used for simulated events and uncertainties on the method used
for instrumental background estimates.  The uncertainties from these sources
are shown in Table~\ref{t-unc}.  These are evaluated by varying each of the
corrections by $\pm 1\sigma$, by comparing different methods (for the
instrumental backgrounds), and by varying the parton distribution functions among
the 20 error sets provided as part of the \cteqSixL\ library.  The variations seen
for different processes for a given uncertainty arise because of differences
among the various background processes and because of intrinsic differences in
the kinematic spectra from different Higgs mass hypotheses.
\begin{table}[htbp]
\caption{\label{t-unc} Systematic uncertainty in background
  and signal predictions given as the fractional uncertainty on the event
  totals.  The ranges correspond to variations introduced by
  different processes (background), the dijet mass window requirement (background
  and signal) and intrinsic differences in kinematics arising from different
  hypothesized Higgs masses (signal).}
\begin{ruledtabular}
\begin{tabular}{ccc} 
 Source                &  Background  & Signal     \\
Lepton ID Efficiencies &  11\% -- 16\% & 11\% -- 12\%  \\
Lepton Resolution      &       2\%    &   2\%      \\
Jet ID Efficiency      &  5\% -- 11\%  &   8\%      \\
Jet Energy Reconstruction  &      10\%    &   7\%      \\
$b$--jet ID Efficiency &  10\% -- 12\% &   9\%      \\
Cross Sections         &   6\% -- 19\% &   7\%      \\
Trigger Efficiency     &      1\%     &   1\%      \\
Instrumental Background&   2\% ($ee$) &            \\
                       &  12\% ($\mu\mu$) &        \\
\end{tabular}
\end{ruledtabular}
\end{table}
\begin{table*}[htbp]
\caption{\label{t-evtcnt} Numbers of predicted background and signal events
  and the observed yield after all selection requirements, including the 
  addition of a dijet mass window.  The mass window is centered on the mean of 
  the reconstructed Higgs mass in simulated $ZH$ events and has a width of $\pm
  1.5\sigma$ in which $\sigma$ is the result of a gaussian fit to the
  reconstructed dijet mass distribution. The upper bounds differ slightly
  between the $\Zee$ and $\Zmm$ events because of different resolutions.  The
  window is applied for illustration, showing the yields in the region of
  highest predicted signal-to-background ratio.  Also shown are the expected
  and observed upper limits on the cross section for the combined analysis at
  95\% confidence level computed as described in the text (without the mass
  window, but weighted by the bin-to-bin signal-to-background ratio).}
\begin{ruledtabular}
\begin{tabular}{lcccccc} 
  & \multicolumn{2}{c}{$M_H = 105$ GeV/$c^2$} &
    \multicolumn{2}{c}{$M_H = 115$ GeV/$c^2$} &
    \multicolumn{2}{c}{$M_H = 125$ GeV/$c^2$}  \\
  & $\ee$ &$ \mm$ & $\ee$ &$ \mm$ & $\ee$ &$ \mm$  \\ \hline
 Mass window(GeV/$c^2$) & $[65,113]$ & $[65,118]$ & $[72,125]$ & $[70,128]$ &
    $[75,136]$ & $[78,137]$ \\
 Predicted signal  & 0.07 & 0.06 & 0.05 & 0.05 & 0.04 & 0.03 \\
 Background        & 1.4  & 3.1  & 1.3  & 3.1  & 1.4  & 2.8  \\
  Data             & 2    & 3    & 1    & 3    & 1    & 4    \\ \hline
 Expected $\sigma_{95}$ & \multicolumn{2}{c}{4.2 pb}
                          & \multicolumn{2}{c}{4.1 pb}
                            & \multicolumn{2}{c}{3.4 pb} \\
 Observed $\sigma_{95}$ & \multicolumn{2}{c}{4.4 pb}
                          & \multicolumn{2}{c}{4.0 pb}
                            & \multicolumn{2}{c}{3.3 pb} \\
\end{tabular}
\end{ruledtabular}
\vskip 2truemm
% \begin{ruledtabular}
% \begin{tabular}{ccccccc} 
%   & \multicolumn{2}{c}{$M_H = 135$ GeV/$c^2$} &
%     \multicolumn{2}{c}{$M_H = 145$ GeV/$c^2$} & 
%     \multicolumn{2}{c}{\hphantom{$M_H = 145$ GeV/$c^2$}}    \\
%   & $\ee$ &$ \mm$ & $\ee$ &$ \mm$ & \hphantom{$\ee$} & \hphantom{$\mm$} \\ \hline
%  Mass Window(GeV/$c^2$) & $[65,113]$ & $[65,118]$ & $[72,125]$ & $[70,128]$ 
%                                   & & \\
%  Predicted Signal  & 0.07 & 0.06 & 0.05 & 0.05 & & \\
%  Background        & 1.4  & 3.1  & 1.3  & 3.1  & & \\
%   Data             & 2    & 3    & 1    & 3    & & \\ \hline
%  Expected $\sigma_{95}$ & \multicolumn{2}{c}{4.2 pb}
%                           & \multicolumn{2}{c}{4.1 pb}
%                           & \multicolumn{2}{c}{}       \\
%  Observed $\sigma_{95}$ & \multicolumn{2}{c}{4.4 pb}
%                           & \multicolumn{2}{c}{4.0 pb}
%                           & \multicolumn{2}{c}{} \\
% \end{tabular}
% \end{ruledtabular}
%\hskip -1.8 truein
\parbox{5.2truein}{
\begin{ruledtabular}
\begin{tabular}{lcccc} 
  & \multicolumn{2}{c}{$M_H = 135$ GeV/$c^2$} &
    \multicolumn{2}{c}{$M_H = 145$ GeV/$c^2$} \\
  & $\ee$ &$ \mm$ & $\ee$ &$ \mm$ \\ \hline
 Mass window(GeV/$c^2$) & $[82,143]$ & $[84,147]$ & $[87,156]$ & $[92,160]$  \\
 Predicted signal  & 0.027& 0.022& 0.015& 0.01 \\
 Background        & 1.6  & 2.9  & 1.6  & 2.8  \\
  Data             & 1    & 5    & 0    & 6    \\ \hline
 Expected $\sigma_{95}$ & \multicolumn{2}{c}{2.8 pb}
                          & \multicolumn{2}{c}{2.6 pb} \\
 Observed $\sigma_{95}$ & \multicolumn{2}{c}{3.1 pb}
                          & \multicolumn{2}{c}{3.4 pb} \\
\end{tabular}
\end{ruledtabular}}
\end{table*}

\begin{figure}[!h]
\includegraphics[scale=0.46]{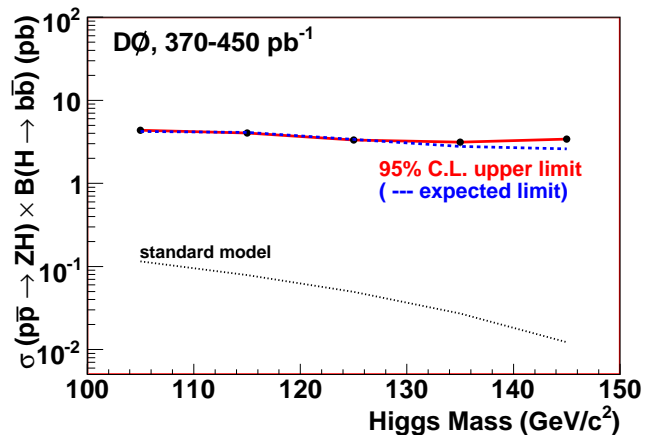}
\caption{\label{f-limit} The expected and observed cross--section limits are
shown as a function of Higgs mass.  The cross section based on the SM is
shown for comparison.}
\end{figure}

The observed yield is consistent with background predictions.  Upper limits on
the $ZH$ production cross section are derived at 95\% confidence level using
the $CL_s$ method~\cite{b-junk}, a modified frequentist procedure, with a
log-likelihood ratio classifier.  The shapes of dijet invariant-mass spectra of
the signal and background are used to produce likelihoods that the data are
consistent with the background-only hypothesis or with a background plus signal
hypothesis. Systematic uncertainties are folded into the likelihoods via
Gaussian distribution, with correlations maintained throughout.  The data
yield, predicted backgrounds and expected and observed limits are shown in
Table~\ref{t-evtcnt} for five hypothetical Higgs masses.  The limits are also
shown in Fig.~\ref{f-limit}.

%95\% confidence
%level upper limits are set on the $ZH$ production cross section using the
%$CL_s$ method~\cite{b-junk}, a modified frequentist procedure, with a log
%likelihood ratio as the test statistic.  The shape of signal and background
%di--$b$--jet invariant--mass spectra are used to produce likelihoods that the
%data are consistent with background only hypothesis or with a background plus
%signal hypothesis, weighted by the predicted signal--to--noise in each bin.
%The procedure also accounts for systematic uncertainties, including
%correlations.

In summary, we have carried out a search for associated $ZH$ production in
events having two high-$p_T$ electrons or muons and two jets identified as
arising from $b$ quarks.  Consistency is found between data and background
predictions.  A 95\% confidence level upper limit on the Higgs
boson cross section $\sigma(\ppbar\rightarrow ZH)\times B(H\rightarrow\bbbar)$
is set between 4.4~pb and 3.1~pb for Higgs bosons with mass between 105~GeV/$c^2$ and
145~GeV/$c^2$, respectively.

% acknowledgement_paragraph_r2.tex                                 4/11/07
%
We thank the staffs at Fermilab and collaborating institutions, 
and acknowledge support from the 
DOE and NSF (USA);
CEA and CNRS/IN2P3 (France);
FASI, Rosatom and RFBR (Russia);
CAPES, CNPq, FAPERJ, FAPESP and FUNDUNESP (Brazil);
DAE and DST (India);
Colciencias (Colombia);
CONACyT (Mexico);
KRF and KOSEF (Korea);
CONICET and UBACyT (Argentina);
FOM (The Netherlands);
PPARC (United Kingdom);
MSMT and GACR (Czech Republic);
CRC Program, CFI, NSERC and WestGrid Project (Canada);
BMBF and DFG (Germany);
SFI (Ireland);
The Swedish Research Council (Sweden);
CAS and CNSF (China);
Alexander von Humboldt Foundation;
and the Marie Curie Program.
%%%% remove Marie Curie at August 07 update
%

%\begin{references}

\end{document}